\documentclass[aps,prl,twocolumn,showpacs,superscriptaddress]{revtex4-2}
\usepackage{graphicx}
\usepackage{amsmath}
\usepackage{amssymb}
\usepackage{colordvi}
\usepackage{mathrsfs}
\usepackage{bm}
\usepackage{verbatim}
\usepackage{dcolumn}
\usepackage{bm}
\usepackage{epsfig}
\usepackage{subfigure}
\usepackage{dsfont}
\usepackage{multirow}
\usepackage[colorlinks,linkcolor=blue,anchorcolor=blue,citecolor=blue]{hyperref}

\begin{document}
\title{Rigid-Body Anisotropy in Noncollinear Antiferromagnets}
\author{Zheng Liu}
\affiliation{CAS Key Laboratory of Strongly-Coupled Quantum Matter Physics, and Department of Physics, University of Science and Technology of China, Hefei, Anhui 230026, China}
\author{Yang Gao}
\email[Correspondence author:~~]{ygao87@ustc.edu.cn}
\affiliation{CAS Key Laboratory of Strongly-Coupled Quantum Matter Physics, and Department of Physics, University of Science and Technology of China, Hefei, Anhui 230026, China}
\affiliation{ICQD, Hefei National Laboratory for Physical Sciences at Microscale, University of Science and Technology of China, Hefei, Anhui 230026, China}
\author{Qian Niu}
\email[Correspondence author:~~]{niuqian@ustc.edu.cn}
\affiliation{CAS Key Laboratory of Strongly-Coupled Quantum Matter Physics, and Department of Physics, University of Science and Technology of China, Hefei, Anhui 230026, China}

\date{\today{}}

\begin{abstract}
Characterizing the anisotropic structure in noncollinear antiferromagnets is essential for antiferromagnetic spintronics. In this work, we provide a microscopic theory linking the anisotropy effects induced by the rigid-body rotation of spin order to spin-orbit coupling.
Our method goes beyond the conventional magnetic group theory, offering a concise yet powerful tool to characterize diverse anisotropy effects in complex magnetic systems.
Using the group representation theory of the spin group, we obtain a set of basis functions formed from tensor elements of spin-orbit vector--which originates from spin-orbit coupling and is tied to the rigid-body rotation of the spin order--to systematically describe the structure of anisotropy effects.
As a concrete example, we apply our framework to coplanar antiferromagnets Mn$_3$Sn and Mn$_3$Ir, demonstrating that the corresponding basis functions can well capture both the geometric and magnitude dependencies of the magnetic anisotropy energy and anomalous Hall conductivity. 
Finally, we discuss the generalization of our framework to broader classes of anisotropy phenomena in magnetic systems.
\end{abstract}

\maketitle

Noncollinear antiferromagnets have drawn significant research interest owing to their unique physical properties and potential for next-generation spintronics\cite{Smejkal2022,Rimmler2025,Baltz2018}.
To realize spintronic applications in noncollinear antiferromagnets, control and detection of the spin order are essential\cite{Jungwirth2018}. 
Recent theoretical and experimental studies have revealed a significant anomalous Hall effect in Mn$_3X$ ($X=$Ga, Ge, Sn, Pt, Ir, Rh) compounds\cite{Chen2014,Kiyohara2016,Zhang2017,Yang2017,Liu2018,Nakatsuji2015,Nayak2016,Chen2021,Ikhlas2022,Pradhan2023}. The Hall response is found to be highly sensitive to the rigid-body rotations of the spin order, providing an electrical probe of the spin order\cite{Nakatsuji2015,Nayak2016,Chen2021,Ikhlas2022,Pradhan2023}. Alongside this, experiments have shown that spin-orbit torques can overcome the magnetic anisotropy barrier, enabling spin-order switching\cite{Tsai2020,Higo2022,Pal2022,Zheng2025,Takeuchi2021,Tsai2021,Krishnaswamy2022,Arpaci2021}.
These advances are intimately linked to spin-orbit coupling mediated spin-lattice interaction, which gives rise to anisotropies in energy, current, and other related observables. Understanding these anisotropy effects is essential for the design of advanced spintronics devices.

However, fundamental challenges remain in characterizing these anisotropy effects at the microscopic level. On one hand, previous studies of anisotropy effects primarily rely on magnetic point groups\cite{Shubnikov1951,Zamorzaev1953,Kleiner1966,BRADLEY1968}, which can determine the existence of anisotropy effects but fail to link their structure to spin order. On the other hand, while spin-orbit coupling breaks spin-group symmetry\cite{ifmmodeSelseSfimejkal2022,Liu2022,Chen2024,Xiao2024,Jiang2024} and is generally considered the origin of anisotropy effects\cite{Kuebler2021,Daalderop1991,Heide2008,Shick2010}, the link between anisotropy effects and spin group symmetry breaking has only been studied in collinear magnets\cite{Liu2025a,McClarty2024,Xiao2025,Roig2025}. A comprehensive and accurate theoretical investigation in more complex noncollinear systems is still lacking. These gaps hinder fundamental studies and device design in noncollinear antiferromagnets, underscoring the urgent need for a deeper theoretical investigation.

In this Letter, we extend our previous framework to noncollinear antiferromagnets, building the explicit connection between various anisotropy effects and spin-orbit coupling. 
By treating spin-orbit coupling perturbatively and parameterizing it via a spin-orbit vector, we develop a group representation theory linking physical observables to this vector. This connection naturally encodes anisotropy effects, as the spin-orbit vector describes the rigid-body rotation of the spin order. 
We show that the unique spin-group symmetries of noncollinear antiferromagnets endow unconventional anisotropic structures.
Taking coplanar antiferromagnets Mn$_3$Sn and Mn$_3$Ir as concrete examples, we derive the forms of magnetic anisotropy energy and anomalous Hall conductivity. In Mn$_3$Sn, the anisotropy energy starts at first order in spin-orbit coupling, in sharp contrast to collinear magnets. Mapping the first-order term to the spin order reveals its intrinsic connection to the  Dzyaloshinskii–Moriya (DM) interaction. We also identify a biaxial single-ion anisotropy energy at second order in spin orbit coupling. In Mn$_3$Ir, capturing the anisotropy of anomalous Hall conductivity requires the inclusion of nonlinear terms in spin-orbit coupling which gives rise to a complex dependence on the spin order.
Our work highlights the central role of spin-orbit coupling in characterizing various anisotropy effects in magnetic systems, thereby paving the way to unveil the intrinsic connection between physical observables and the spin order.

\begin{table*}
	\caption{The irreducible representations of different physical bases and corresponding spin-orbit vectors.}
	\begin{ruledtabular}
		\begin{tabular}{cccc}
			Quantity                  &Irreducible representations                          & Physical bases                              & Spin-orbit vectors  \\ \hline
			\multirow{7}{*} {Scalar}  &\multirow{7}{*} {$A_g\otimes A_{1g}$}        & \multirow{7}{*} {$E$}           & $O_3^3$                                                   \\  
			&                                             &                                 & $O_3^3O_3^3$                                              \\          
			&                                             &                                 & $O_3^1O_3^1+O_3^2O_3^2$                        \\
			&                                             &                                 & $O_1^3O_1^3+O_2^3O_2^3$                        \\
			&                                             &                                 & $-O_1^2O_2^1+O_1^1O_2^2$                       \\
			&                                             &                                 & $O_1^1O_1^1-2O_1^2O_2^1+O_2^2O_2^2$            \\
			&                                             &                                 & $O_1^2O_1^2+2O_1^2O_2^1+O_2^1O_2^1$            \\
			\hline
			Pseudovector                                  & $A_u\otimes T_{1g}$             & $(\sigma_x^H,\sigma_y^H,\sigma_z^H)$       & $(-2O_1^1+O_1^2+O_1^3,O_2^1-2O_2^2+O_2^3,O_3^1+O_3^2-2O_3^3) $
		\end{tabular}
	\end{ruledtabular}
	\label{table-1}
\end{table*}

{\it Spin group analysis.---} In noncollinear antiferromagnets, strong exchange coupling stabilizes the spin texture, allowing rigid-body rotation of the spin order as a low-energy, relatively accessible degree of freedom. This collective rotation of the spin order is governed by SO(3) group symmetry, which includes all proper rotations and can be fully described by three Euler angles.

Generally, spin-orbit coupling is the source of anisotropy effects under the rigid-body rotation of the spin order. To see this, let us first consider the single-particle Hamiltonian without spin-orbit coupling, which can be expressed as
\begin{eqnarray}
	\hat{H}_0=\frac{\hat{\boldsymbol{p}}^2}{2m}+V(\boldsymbol{r})+\boldsymbol{m}(\boldsymbol{r})\cdot\hat{\boldsymbol{\sigma}},
\end{eqnarray}
where $\bm m(\bm r)$ represents the $\bm r$-dependent exchange field that captures the noncollinear magnetic configuration. In the absence of spin-orbit coupling, the lattice and spin coordinate systems exhibit independent gauge freedoms. By fixing the lattice frame and allowing the electron spin frame to rigidly rotate with the spin order, the Hamiltonian $\hat{H}_0$ remains invariant, rendering all spin-independent physical quantities isotropic. Hence, spin-orbit coupling is indispensable for anisotropy effects. 

Based on the above definition of the spin frame, the spin-orbit coupling takes the following form
\begin{eqnarray}\label{eq:Ham_soc2}
	\hat{H}_{\mathrm{SOC}}=\frac{\hbar}{m^2c^2}\sum_{ij}O_i^j\hat{L}_i\hat{\sigma}_j,
\end{eqnarray}
where $\hat{\bm L}=\boldsymbol{\nabla}U(\bm r)\times \hat{\bm p}$. The misalignment between the lattice frame and the spin frame is characterized by the spin-orbit vector $O_i^j$ with $i$ and $j$ in lattice and spin space, respectively. Specifically, the orientation of the $j$-th spin axis in the lattice frame is described by the vector $\bm{O}^j=(O_1^j,O_2^j,O_3^j)$, which transforms as $\bm O^j=\boldsymbol{\mathcal{R}}\cdot \bm O^{j,0}$. Here, $\boldsymbol{\mathcal{R}}$ denotes the rotation matrix corresponding to the rigid-body rotation of spin order, while $\bm O^{j,0}=(\delta_{1j},\delta_{2j},\delta_{3j})$ represents the initial orientation of the $j$-th spin axis. This directly leads to $O_i^j=\mathcal{R}_{ij}$, demonstrating how the global rotation of the spin order is encoded in the spin-orbit vector.

It is apparent that $\hat{H}_{\mathrm{SOC}}$ varies under rigid-body rotation of the spin order, indicating its intimate connection with anisotropy effects. As the spin-orbit coupling is usually small in strong magnets\cite{Cowan1981,Stoehr2006,Novak2001,Dunn1961,Yuan2017,Stamokostas2018,Tanaka2008,Naito2010,Herman1963}, it can be treated as a perturbation. Anisotropy effects can then be expanded in powers of the spin-orbit vector under the symmetry constraints of $\hat{H}_0$. Therefore, the anisotropic structure of diverse physical quantities can be systematically characterized within the SO(3) space spanned by Euler-angle rotations of the spin order.

To obtain the expansion of anisotropy effects, we first examine the symmetry constraints of $\hat{H}_0$. The symmetry properties of $\hat{H}_0$ are analyzed using the spin-group formalism\cite{ifmmodeSelseSfimejkal2022,Liu2022,Xiao2024,Jiang2024,Chen2024,Yang2024,Chen2025}. Crucially, in the previously defined spin frame, $\hat{H}_0$ remains invariant under rigid-body rotation of spin order, as does the associated spin group. The typical group element $\{C_{\boldsymbol{m}}^s(\phi)|C_{\boldsymbol{n}}^L(\theta)\}$ contains two parts of symmetry operations, with the left operation acting exclusively in the spin space and the right operation solely in lattice space. This differs fundamentally from collinear ferromagnets, where spin-space and lattice-space operations are completely decoupled\cite{Liu2025a}.  
Under spin-group operations, $\boldsymbol{O}$ transforms as $O_i^j\rightarrow R_{ik}S_{jl}O_k^l$, where $\bm R$ and $\bm S$ are proper rotation components in lattice space and spin space, respectively. Note that time-reversal symmetry simultaneously flips the signs of $\hat{\bm L}$ and $\hat{\bm \sigma}$ and thus leaves the spin-orbit vector unaffected.

The expansion can then be analyzed by utilizing representation theory. Any physical observable $\bm{F}$ can be decomposed into one or more components $\boldsymbol{\mathcal{F}}^i$ that transform according to the $i$-th irreducible representation of the spin group. Consequently, $\boldsymbol{\mathcal{F}}^i$ can be connected to the spin-orbit vector by
\begin{align}\label{eq:Any_expand2}
	\boldsymbol{\mathcal{F}}^i=\sum_{n,k}c_{nk} \bm f^{i}_{nk}(\boldsymbol{O}),
\end{align}
where $c_{nk}$ is a constant coefficient, and $\bm f^i_{nk}(\boldsymbol{O})$ denotes the basis functions, belonging to the same irreducible representations of $\boldsymbol{\mathcal{F}}^i$, at the $n$-th order of spin-orbit vector. The index $k$ labels symmetry-equivalent basis functions.
Notably, the $n$-th order term's contribution is proportional to the $n$-th power of the spin-orbit coupling. Given that spin-orbit coupling is relatively weak compared to the typical energy scale in $\hat{H}_0$, the anisotropy strength will decay exponentially with increasing spin-orbit vector order.

The core idea of this framework is to see the spin group as a ``ground" and spin-orbit coupling as a perturbation for the expansion of anisotropy effects, i.e., viewing anisotropy effects as spin-group symmetry breaking phenomena. Different from the analyses based on magnetic point groups\cite{Shubnikov1951,Zamorzaev1953,Kleiner1966,BRADLEY1968}, which mainly confirm the presence of anisotropy effects, our theory provides detailed structure information about these anisotropy effects. Our framework also differs fundamentally from cluster multipole theory\cite{Suzuki2017,Suzuki2019,Bhowal2024,Yatsushiro2021,Watanabe2018,Hayami2018}. In the latter, real-space magnetic multipoles serve as point-group-symmetry-breaking terms that are linear in the spin order and are associated with the exchange field energy scale. In contrast, our theory treats spin-orbit coupling as the symmetry-breaking perturbation, which is generally the smallest energy scale in strong magnets and directly governs magnetic anisotropy. This allows us to establish a quantitative connection between anisotropy effects and spin-orbit coupling, while also enabling a natural extension to nonlinear terms in the spin order.

{\it Energy magnetic anisotropy.---} Armed with this theoretical framework, we then analyze the magnetic anisotropy energy, which is a prototypical manifestation of spin-orbit coupling\cite{Kuebler2021,Daalderop1991}. As a scalar, the anisotropy energy belongs to the trivial irreducible representation of the spin group. Consequently, establishing its connection to spin-orbit coupling requires identifying polynomials of the spin-orbit vector that also transform as the trivial representation.

\begin{figure}
	\includegraphics[width=8.5cm,angle=0]{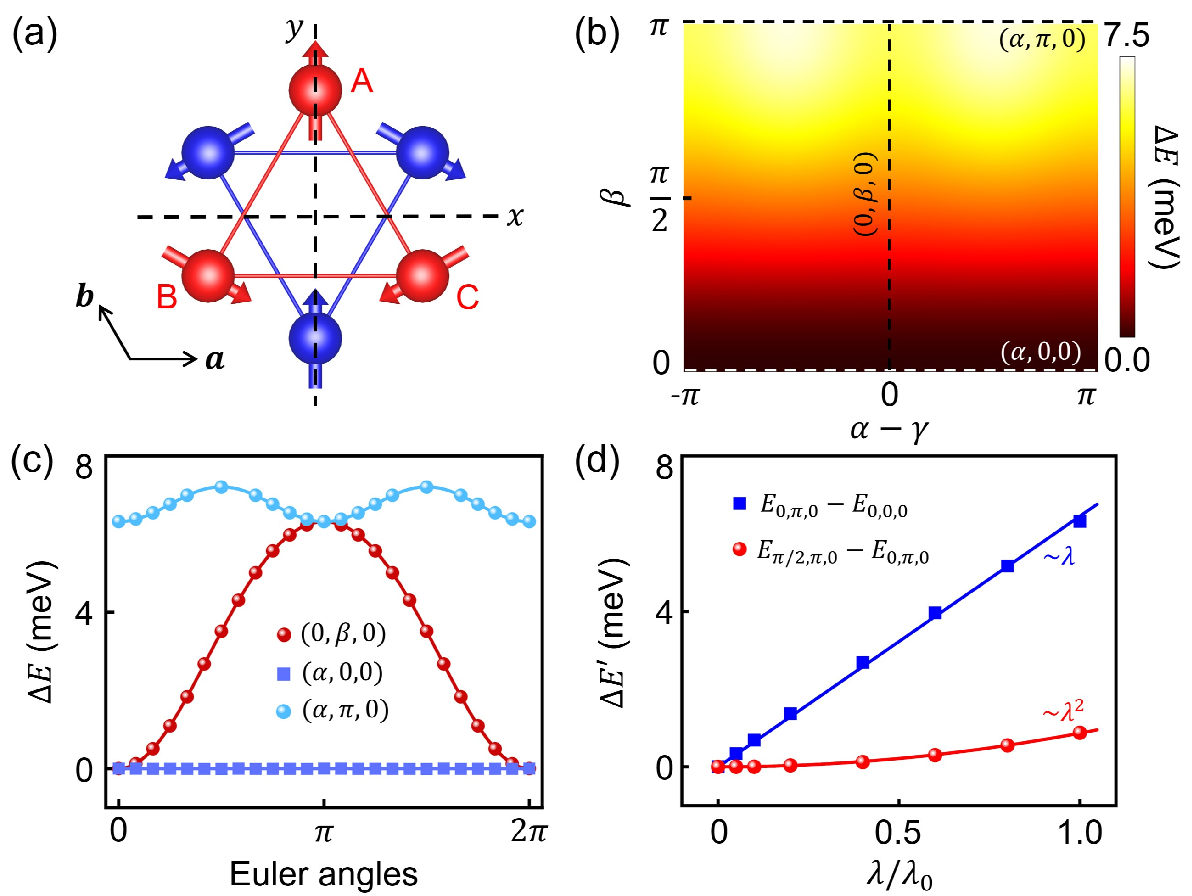}
	\caption{Magnetic properties and magnetic anisotropy energy of Mn$_3$Sn. (a) Magnetic configuration. (b) Anisotropy energy in the Euler angle space obtained from Eq.~\ref{eq:MAE_Mn3Sn}. (c) Magnetic anisotropy energy $\Delta E$ as a function of Euler angles $(\alpha,0,0)$, $(0,\beta,0)$, and $(\alpha,\pi,0)$. Points and solid lines represent calculated and fitted results, respectively. (d) Energy difference of $E_{0,\pi,0}-E_{0,0,0}$ and $E_{\pi/2,\pi,0}-E_{0,\pi,0}$ as a function of the coupling strength, where $\lambda_0$ represents the actual strength of spin-orbit coupling.}
	\label{Fig2}
\end{figure}

Without loss of generality, we consider the magnetic anisotropy energy in the noncollinear antiferromagnet Mn$_3$Sn. 
The magnetic configuration of Mn$_3$Sn, illustrated in Fig.~\ref{Fig2}(a), forms a coplanar triangular antiferromagnetic structure\cite{Chen2021,Nakatsuji2015}. Its spin point group can be expressed as a direct product of a spin-only group and a nontrivial spin group, which is $G_{SP}=Z_2^K\otimes G_{NSP}$. Here, $Z_2^K$ consists of two elements: $E^s$ and $I^sC_{2z}^s$. $G_{NSP}$ is generated by $P$=$\{E^s|I^L\}$, $C_{6z}=\{C_{3z}^s|C_{6z}^L\}$, and $C_{2x}=\{C_{2y}^s|C_{2x}^L\}$, where $I$ denotes inversion symmetry. The spin-only group $Z_2^K$ is isomorphic to $S_2$, while the nontrivial spin group $G_{NSP}$ is isomorphic to $D_{6h}$. The corresponding group elements and character tables are provided in Supplemental Material\cite{Supplemental}.

We then connect the magnetic anisotropy energy and spin-orbit vector. The magnetic anisotropy energy is defined as $\Delta E=E_{\alpha,\beta,\gamma}-E_{0,0,0}$, where $E_{\alpha,\beta,\gamma}$ represents the total energy of Mn$_3$Sn after a rigid-body rotation of magnetization by the Euler angles $(\alpha,\beta,\gamma)$, defined in the $Z_{\alpha}Y_{\beta}Z_{\gamma}$ convention, and $E_{0,0,0}$ is the total energy of the initial magnetic configuration shown in Fig.~\ref{Fig2}(a). As mentioned above, the total energy belongs to the trivial irreducible representation which is $A_g\otimes A_{1g}$ in Mn$_3$Sn. In Table~\ref{table-1}, we present the corresponding basis functions up to the second order of the spin-orbit vector. 

Remarkably, a first-order term in the spin-orbit vector emerges in the magnetic anisotropy energy, as shown in Table~\ref{table-1}. This is in sharp contrast to collinear magnets. In collinear systems, the spin-only group must include the generators $C_{\infty z}^s$ and $I^sC_{2x}^s$, where $z$ denotes the magnetic moment direction\cite{Liu2022}. These group elements forbid the appearance of the trivial representation in the first-order expansion of spin-orbit vector, conclusively demonstrating the distinct feature of noncollinear antiferromagnets. 

The expression for the magnetic anisotropy energy can be obtained by using Eq.~\ref{eq:Any_expand2}. Under the rigid-body rotation of Euler angles $(\alpha,\beta,\gamma)$, the spin-orbit vector has the form $\bm{O}=R_z(\alpha)R_y(\beta)R_z(\gamma)$,
where $R_i(\theta)$ denotes the rotation matrix about axis $i$ through angle $\theta$. The first-order anisotropy energy term is $\Delta E^1\sim 1-\cos\beta$, where $\beta$ denotes the tilt of magnetic moments out of the $xy$-plane. 
Summing over the basis functions in Table~\ref{table-1}, we obtain the following form
\begin{align}\label{eq:MAE_Mn3Sn}
	\Delta E=a-a \cos\beta+b\sin^2\beta+c\sin^4\frac{\beta}{2}\sin^2(\alpha-\gamma),
\end{align}
with $a$, $b$, and $c$ being the coefficients.

Figure~\ref{Fig2}(b) shows the anisotropy energy in magnetization space, calculated using Eq.~\ref{eq:MAE_Mn3Sn}. The fitting yields coefficients $a=3.159$, $b=0.351$, and $c=0.890$ in units of meV, based on anisotropy energies at three representative Euler angles: $(\alpha,0,0)$, $(0,\beta,0)$ and $(\alpha,\pi,0)$, as illustrated by dashed lines in Fig.~\ref{Fig2}(b). The corresponding data points are shown in Fig.~\ref{Fig2}(c), where good agreement with the fitted curves confirms the validity of our theory. 
As shown in Fig.~\ref{Fig2}(b), the anisotropy energy increases significantly with $\beta$, indicating a significant barrier for spins rotating out of the $xy$-plane. In addition, $\Delta E$ varies clearly with the $\alpha-\gamma$ angle difference, as evidenced by the color variation from $-\pi$ to $\pi$.

To clarify the link between the anisotropy energy and the spin-orbit vectors, we further examine $\Delta E$ along three distinct directions. The calculated $\Delta E$ at Euler angle $(0,\beta,0)$ is shown in Fig.~\ref{Fig2}(c) (red dots). A significant energy barrier appears as $\beta$ varies from $0$ to $\pi$, which primarily originates from the first-order term $O_3^3$ with a contribution $\Delta E^1\sim 1-\cos\beta$. To verify this, Fig.~\ref{Fig2}(d) plots the energy difference $E_{0,\pi,0}-E_{0,0,0}$ as a function of the spin-orbit coupling strength $\lambda$. The first-principles calculated results scale linearly with $\lambda$, confirming that the barrier is dominated by the first-order term.

We further investigate the origin of the linear term $O_3^3$ and its physical significance. By mapping the spin-orbit vector to spin order\cite{Supplemental}, we recognized that $O_3^3$ is connected to the in-plane DM interaction term $D\sum_{ij}\hat{\bm{z}}\cdot (\bm{S}_i\times\bm{S}_j)$. This term corresponds to a large energy barrier for moments rotating out of the $xy$-plane, stabilizing the in-plane configuration of the spin order. This result is also consistent with the well-known fact that the DM interaction generally emerges at the first order of spin–orbit coupling\cite{Moriya1960,Wang2023}.

The anisotropy energies at Euler angles $(\alpha,\pi,0)$ and $(\alpha,0,0)$ are shown in Fig.~\ref{Fig2}(c). Here, $\beta$ is fixed, so the variation of $\Delta E$ arises solely from second-order terms in spin-orbit vectors. At $(\alpha,\pi,0)$, $\Delta E$ follows $\Delta E\sim \sin^2\alpha$, in excellent agreement with the numerical results in Fig.~\ref{Fig2}(c). The energy difference $E_{\pi/2,\pi,0}-E_{0,\pi,0}$ as a function of spin-orbit coupling strength is illustrated in Fig.~\ref{Fig2}(d). The calculated values exhibit a clear quadratic dependence on $\lambda$, confirming the dominant role of the second-order terms. At $(\alpha,0,0)$, our theory predicts $\Delta E=0$, in good agreement with the calculated results in Fig.~\ref{Fig2}(c). This indicates that the spin order can rotate freely within the $xy$-plane, consistent with the experimental observations\cite{Tomiyoshi1982}. We further analyze the anisotropy energy as a function of the spin orientation\cite{Supplemental} and find that the biaxial single-ion anisotropy term ensures the spin energy remains unchanged under rigid-body rotation of the spin order within the $xy$-plane. This result is fully consistent with our spin-orbit coupling framework.

Altogether, this analytical form of the anisotropy energy provides a foundation for identifying the magnetic ground-state configuration, as well as for exploring magnetic dynamics and controlling spin textures. Our results uncover fundamental insights into the interplay between spin-orbit coupling and anisotropy energy, and extend naturally to other noncollinear magnets with distinct spin group symmetries\cite{Supplemental}.

{\it Anomalous Hall magnetic anisotropy.---} In addition to the magnetic anisotropy energy, the anisotropy property of the anomalous Hall effect originates from the spin-group symmetry breaking, which is also a spin-orbit coupling effect\cite{Karplus1954,Nagaosa2010}. As a lattice-space pseudovector, the anomalous Hall conductivity will couple to basis functions of spin-orbit vector sharing the same pseudovector symmetry.

\begin{figure}
	\includegraphics[width=8.5cm,angle=0]{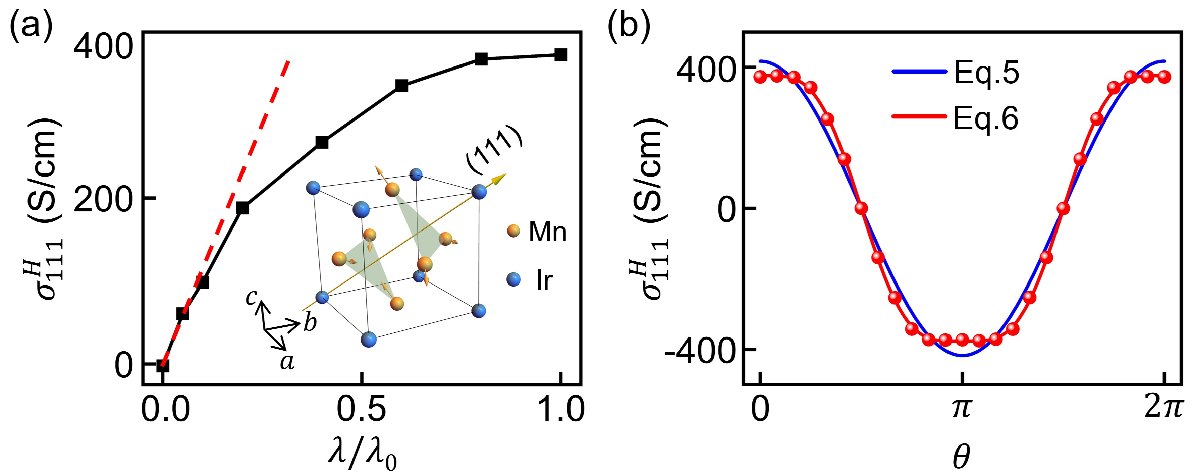}
	\caption{Anomalous Hall effect of Mn$_3$Ir. (a) Anomalous Hall conductivity as a function of spin-orbit coupling strength, with $\lambda_0$ denoting the intrinsic value. The inset shows the lattice structure and magnetic configuration of Mn$_3$Ir. (b) Anomalous Hall conductivity $\sigma_{111}^H$ versus rotation angle $\theta$ about  the [111] axis. Points and solid lines represent calculated and fitted results, respectively.}
	\label{Fig_Mn3Ir}
\end{figure}

We focus on the anomalous Hall effect in Mn$_3$Ir\cite{Chen2014}. The lattice structure and energetically stable spin configuration of Mn$_3$Ir are shown in the inset of Fig.~\ref{Fig_Mn3Ir}(a). Its spin group is the direct product of the spin-only group ($Z_2^K$) and nontrivial spin group ($G_{NSP}$). The spin-only group $Z_2^K$, isomorphic to the $S_2$ point group, comprises two spin-space operations: $E^s$ and $I^sC_{2(111)}^s$. Remarkably, $G_{NSP}$ is isomorphic to the $O_h$ point group, with its elements classified into ten distinct conjugacy classes\cite{Supplemental}.

The anomalous Hall conductivity $\sigma_{111}^H$ as a function of spin-orbit coupling strength $\lambda$ is illustrated in Fig.~\ref{Fig_Mn3Ir}(a) by black squares. The conductivity $\sigma_{111}^H$ deviates significantly from linear behavior as $\lambda$ increases, indicating the important role of nonlinear spin-orbit coupling terms in determining the anomalous Hall conductivity. For comparison, we perform analogous calculations for Mn$_3$Sn, where $\boldsymbol{\sigma}^H$ varies linearly with $\lambda$, and find that the anomalous Hall conductivity in Mn$_3$Sn can be well-described by first-order spin-orbit coupling terms\cite{Supplemental}.

For conciseness, we focus on the anomalous Hall conductivity of Mn$_3$Ir in the main text. $\boldsymbol{\sigma}^H$ belongs to the $A_u\otimes T_{1g}$ irreducible representation of the spin group. The corresponding spin-orbit vectors, which share the same representation as $\boldsymbol{\sigma}^H$ to first order are illustrated in Table~\ref{table-1}.
Under a rotation of the spin order by an angle $\theta$ about the [111] direction, the spin-orbit vector has the form $\bm O=R_{111}(\theta)$. Using Eq.~\ref{eq:Any_expand2}, we find that the anomalous Hall conductivity $\boldsymbol{\sigma}^H$ remains always aligned with the [111] direction, with its magnitude following
\begin{eqnarray}\label{eq:sigma_Mn3Ir}
	\sigma_{111}^H=\alpha_0\cos\theta,
\end{eqnarray}
where $\alpha_0$ is a constant. 
We plot the first-principles calculated $\sigma_{111}^H$ as red dots in Fig.~\ref{Fig_Mn3Ir}(b). One can find that Eq.~\ref{eq:sigma_Mn3Ir} (blue curve) captures the overall trend of $\boldsymbol{\sigma}^H$, but significant deviations persist near $0$, $\pi$, and $2\pi$. These deviations come from nonlinear orders of spin-orbit coupling that correspond to nonlinear terms in the spin order.

To account for the higher-order corrections to $\boldsymbol{\sigma}^H$, nonlinear terms in the spin-orbit vector are further included. The corresponding basis functions up to the second and third orders are provided in Supplemental Material\cite{Supplemental}. Although the number of symmetry-equivalent basis functions is enormous, the undetermined coefficients under rotation about the [111] axis exhibit a remarkably simple structure. The second-order term retains a $\cos\theta$ dependence, while the third-order term introduces a new angular pattern, $\sigma_{111}^H\propto\cos\theta\cos2\theta$. Accordingly, the expression up to the third order in spin-orbit coupling takes the form
\begin{eqnarray}\label{eq:Mn3Ir_3order}
	\sigma_{111}^H=\alpha_0\cos\theta+\beta_0\cos\theta\cos2\theta,
\end{eqnarray}
where $\alpha_0$ and $\beta_0$ are coefficients. One can find that Eq.~\ref{eq:Mn3Ir_3order} (red curve) fits the calculated data well with the coefficients $\alpha_0=473.5$ and $\beta_0=-96.3$ in units of S/cm, demonstrating the predictive power of our theory. 
This anisotropic structure can serve as a sensitive probe for spin texture detection.

Finally, we emphasize that spin-orbit coupling is indispensable for the anomalous Hall effect in coplanar antiferromagnets like Mn$_3$Sn and Mn$_3$Ir, where $I^sC_{2\bm n}^s$ spin-group symmetry forbids a scalar representation of the Hall conductivity. Thus, $\boldsymbol{\sigma}^H$ must vanish in the absence of spin-orbit coupling.  In contrast, noncoplanar antiferromagnets such as CoNb$_3$S$_6$ and CoTa$_3$S$_6$\cite{Ghimire2018,Takagi2023,Park2023} exhibit a component of $\boldsymbol{\sigma}^H$ that transforms as a scalar under the spin group, allowing a finite Hall conductivity at zeroth order in spin-orbit coupling. This spin-orbit-coupling-independent contribution, arising from the noncoplanar spin texture, remains invariant under rigid-body rotations of the spin order. However, the anisotropy effects of anomalous Hall effect still require spin-orbit coupling to break the spin-group symmetry.

{\it Summary.---} In this work, we establish a group representation theory that describes anisotropy effects more accurately in noncollinear antiferromagnets. 
Using coplanar antiferromagnets Mn$_3$Sn and Mn$_3$Ir as examples, we demonstrate that our theory can faithfully reflect the structure and magnitude of the anisotropy energy and the anomalous Hall effect.
Our theory is applicable to a broad class of magnetic systems, including ferromagnets and altermagnets\cite{Mazin2022,ifmmodeSelseSfimejkal2022a,Liu2025}, as well as to other phenomena such as anisotropic magnetoresistance\cite{Ritzinger2023a}, the anomalous Nernst effect\cite{Ikhlas2017}, the nonlinear Hall effect\cite{Shao2020,Wang2021,Liu2021,Wang2023a}, and the spin-orbit-coupling-induced magnetism\cite{Liu2025b}.

\begin{acknowledgments}
The authors are supported by the National Natural Science Foundation of China (12234017). Y. G. is also supported by the National Natural Science Foundation of China (12374164). Z. L. is also supported by Postdoctoral Fellowship Program of CPSF (GZC20232562) and fellowship from the China Postdoctoral Science Foundation (2024M753080). Y. G. and Q. N. are also supported by the Innovation Program for Quantum Science and Technology (2021ZD0302802). The supercomputing service of USTC is gratefully acknowledged.
\end{acknowledgments}

\end{document}